\documentclass[letterpaper,10 pt,conference]{ieeeconf}
\IEEEoverridecommandlockouts 
\usepackage{amsmath}
\usepackage{algorithmic}
\usepackage{array}
\usepackage{stfloats}
\usepackage{url}
\usepackage{cite}
\usepackage{graphicx}
\usepackage{makecell}
\usepackage{mathtools}
\usepackage{threeparttable}
\usepackage{amssymb}
\usepackage{bm}
\usepackage{array} 
\usepackage{float}
\usepackage{subfigure}
\usepackage{gensymb}
\usepackage{textcomp}
\usepackage{wasysym}
\usepackage[table,dvipsnames]{xcolor}
\usepackage{hyperref}
\usepackage{booktabs}
\usepackage{ragged2e}
\usepackage{blkarray,bigstrut}
\usepackage{multirow}
\usepackage{colortbl,hhline}
\usepackage{hyphsubst}
\usepackage{makecell}
\HyphSubstLet{english}{usenglishmax}
\usepackage[english]{babel}
\usepackage{color,soul}
\usepackage{etoolbox}
\makeatletter
\patchcmd{\@makecaption}
  {\scshape}
  {}
  {}
  {}
\makeatother

\title{\LARGE \bf
Domain Adaptive Sim-to-Real Segmentation of Oropharyngeal Organs Towards Robot-assisted Intubation
}

\author{Guankun Wang, Tian-Ao Ren, Jiewen Lai, Long Bai, and Hongliang Ren, \emph{Senior Member, IEEE}% <-this % stops a space
\thanks{This work was supported by HK RGC CRF-C4026-21GF. G. Wang, J. Lai, L. Bai, and H. Ren are with CUHK, Hong Kong, China. T.-A. Ren is with BUCT, Beijing, China. (\textit{Corresponding Author: H. Ren})
    }
}

\begin{document}

\maketitle
\thispagestyle{empty}
\pagestyle{empty}

%%%%%%%%%%%%%%%%%%%%%%%%%%%%%%%%%%%%%%%%%%%%%%%%%%%%%%%%%%%%%%%%%%%%%%%%%%%%%%%%
\begin{abstract}
Robotic-assisted tracheal intubation requires the robot to distinguish anatomical features like an experienced physician using deep-learning techniques. However, real datasets of oropharyngeal organs are limited due to patient privacy issues, making it challenging to train deep-learning models for accurate image segmentation. We hereby consider generating a new data modality through a virtual environment to assist the training process. Specifically, this work introduces a virtual dataset generated by the Simulation Open Framework Architecture (SOFA) framework to overcome the limited availability of actual endoscopic images. We also propose a domain adaptive Sim-to-Real method for oropharyngeal organ image segmentation, which employs an image blending strategy called IoU-Ranking Blend (IRB) and style-transfer techniques to address discrepancies between datasets. Experimental results demonstrate the superior performance of the proposed approach with domain adaptive models, improving segmentation accuracy and training stability. In the practical application, the trained segmentation model holds great promise for robot-assisted intubation surgery and intelligent surgical navigation.
\end{abstract}

\section{Introduction}\label{sec1}

Transoral tracheal intubation (TI) is the gold standard for securing a patient's airway when they require respiratory assistance. However, the success of this procedure hinges on the physician's skills to correctly insert an endotracheal tube into the patient's trachea \cite{thomas2014tracheal}. On top of video-assisted transoral TI, robot-assisted transoral TI makes intubation even more effective by automating the process, but requires accurate segmentation of anatomical features to function properly \cite{caplan2003practice}. This motivates the work herewith.

\begin{figure}[t!]
    \centering
    \includegraphics[width=3.2in, trim=0 45 0 0]{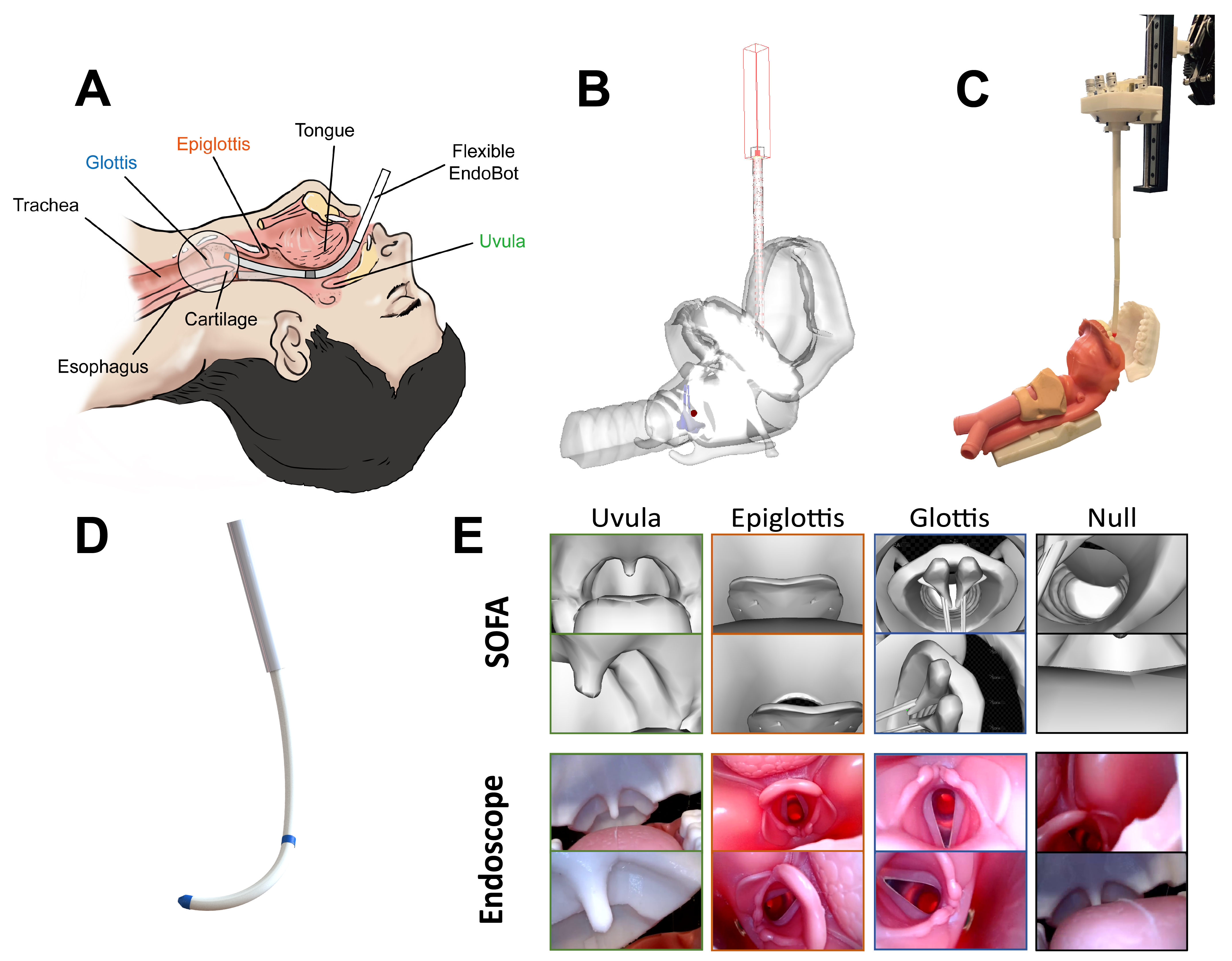}
    \caption{(A) Using a flexible robotic endoscope (EndoBot) as a stylet to guide the next step's tracheal intubation. (B) SOFA scene. (C) Real-world scene. (D) CAD design of our EndoBot (adapted from~\cite{lai2022constrained}) that was used in the virtual and reality setup. (E) Dataset examples.}
    \label{Figure2}
\end{figure}
As shown in Fig.~\ref{Figure2}-A, robot-assisted TI employs a steerable flexible endoscope that works as a style to navigate to the repository tract instead of the digestive tract with the aid of endoscopic vision. Segmenting oropharyngeal organs is a critical step in robot-assisted TI. However, it is challenging to obtain real datasets of oropharyngeal organs due to patient privacy and the need for sufficient data. %\cite{lee2020clinical} 
Deploying supervised learning techniques for segmentation tasks with insufficient data would result in poor performance. To alleviate this problem, synthetic or simulation data for Sim-to-Real transfer have been exploited ~\cite{frangi2018simulation}. Our work employs Simulation Open Framework Architecture (SOFA)\footnote{\href{https://www.sofa-framework.org/}{https://www.sofa-framework.org/}}-generated scenes as the data source, as a lower-quality data source is ideal for testing the domain adaptiveness in Sim-to-Real deployment. Since the oropharyngeal organs in simulation images are constrained in representing natural textures like color and reflection, segmentation performance will be significantly degraded in real datasets (domains) if the model training relies on simulated datasets only, making the Sim-to-Real deployment challenging. 

%    \caption{Overview of the proposed domain adaption training. A subset of the images from the target domain is first blended into the source domain. Then, the style-transfer is used to further optimize the source domain images. The new source domain is used as the training set for segmentation.}
\begin{figure}
    \centering
    \includegraphics[width=3.0in, trim=0 35 0 0]{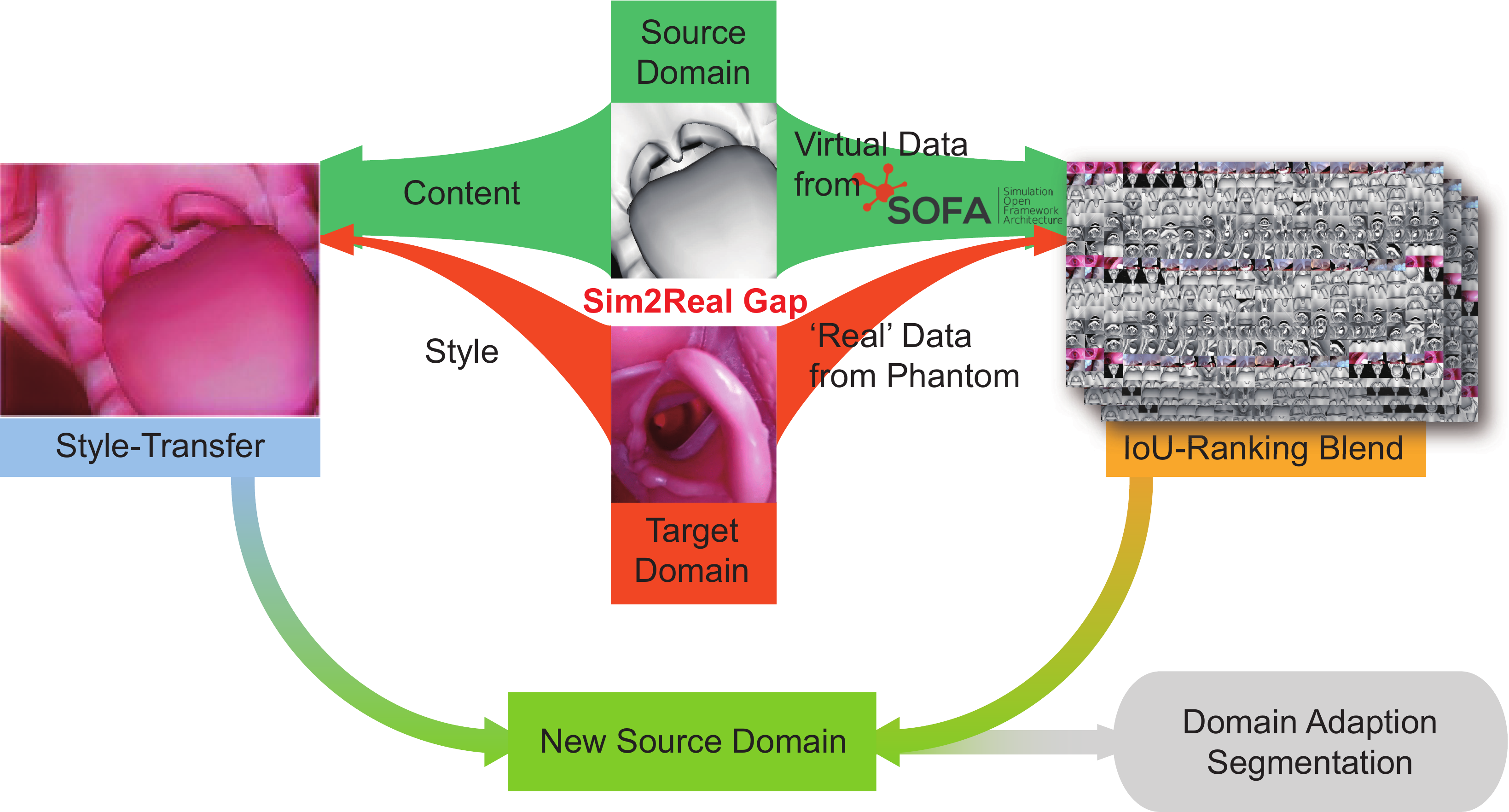}
    \caption{Overview of the proposed domain adaption training. The source domain is first reconstructed based on the IRB strategy. Then, the style-transfer module narrows the Sim-to-Real gap between the two domains. The new source domain will be used as the training set for segmentation.}
    \label{Figure1}
    \vspace{-1em}
\end{figure}

Given the problems mentioned above, reducing the differences between datasets is the most direct and effective way. In this work, we apply a semi-supervised learning method with a novel image-blending (IRB) strategy. We further apply the IRB method with style-transfer techniques to the domain adaption segmentation of oropharyngeal organs. Firstly, we use the SOFA framework to reconstruct a virtual scene and generate the source domain (Fig. \ref{Figure2}-B). Furthermore, real images are captured on a real-world phantom, which is used as the target domain (Fig.~\ref{Figure2}-C). Some sample images are demonstrated in Fig.~\ref{Figure2}-E. Then, to improve the performance of segmentation models in the Sim-to-Real task, we try to blend a small batch of real images into the simulation domain with the Intersection over Union (IoU)-Ranking Blend (IRB) mechanism. The mechanism sorts the resultant IoU among classes after training and refines the blending proportion for the next training iteration according to the sorting. In this regard, the potential of a limited number of mixed images can be fully utilized in the training process to improve the segmentation performance of real domains. Finally, the style-transfer technique is used to reduce the differences between the source domain and the target domain from the image style. The proposed framework is depicted in Fig.~\ref{Figure1}.

\section{Domain Adaptive Sim-to-Real with IRB}
\label{domain adaptive}

\begin{figure}[t!]
    \centering
    \includegraphics[width=3.2in, trim=0 35 0 0]{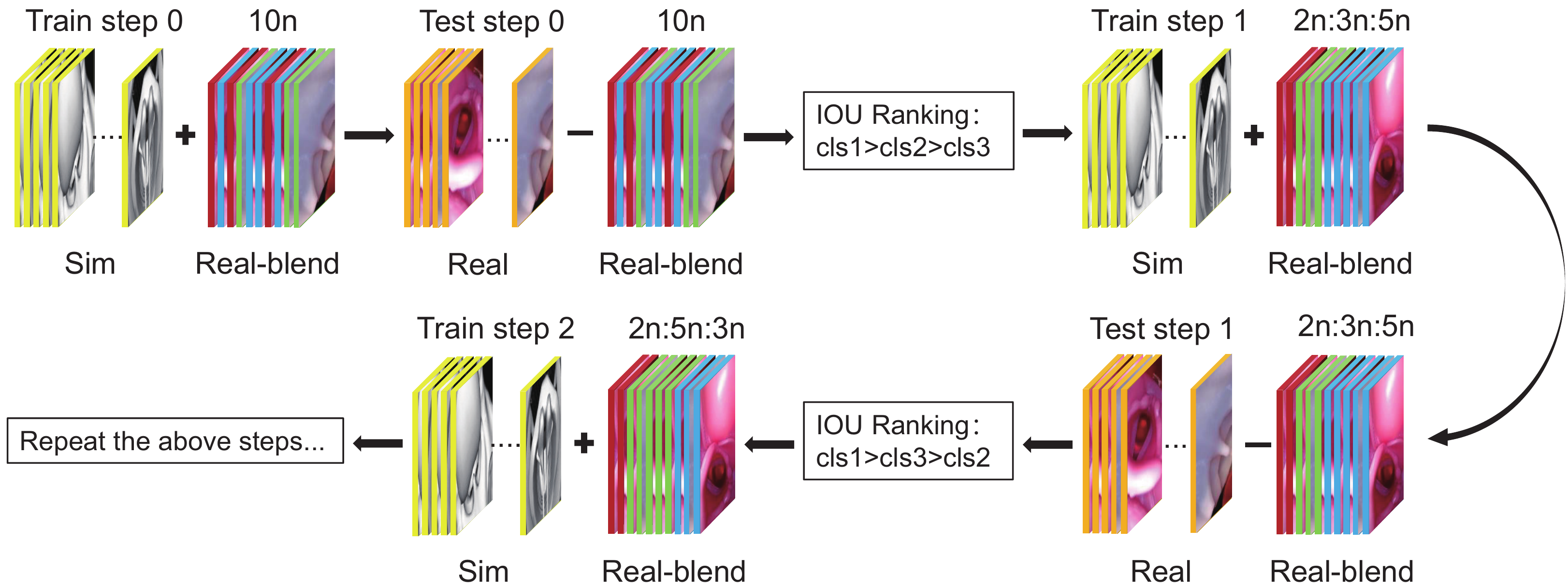}
    \caption{Flow chart of the IRB strategy. \emph{10n} and \emph{cls} represent the number of blended images in the class, respectively. The colors of three \emph{cls} are red, green and blue. The above steps will be repeated until there is no new IoU ranking. The best mIoU will be selected from all test results.}
    \label{Figure3}
    \vspace{-0.5em}
\end{figure}

In this work, the IRB method mixes images based on each class's segmentation testing results after training the model, which can increase the content similarity between datasets. Since there are three vital oropharyngeal organs (uvula, epiglottis, and glottis), a ratio of 5:3:2 for three classes is applied to investigate the IRB strategy, while the number of blends is 10 to 40. Fig.~\ref{Figure3} shows the flow chart of the blending sequence. 
Style mainly refers to the textures, colors, and visual patterns in images at various spatial scales, and it is the low-level image features. The style-transfer technique provides a new viewpoint to narrow the differences between the source and target domain in terms of image style. Moreover, the image content, which is the high-level features, is optimized by the IRB method in the previous section. Consequently, both jointly modify the distribution between datasets and improve the performance of the segmentation model from the perspective of the low-level and high-level features of the images, respectively.

\section{Experiments} \label{experiment}

We conduct extensive experiments to demonstrate the effectiveness of IRB and the style-transfer technique on domain adaptation segmentation. We make the comparison before and after the introduction of our proposed method on state-of-the-art (SOTA) domain adaptive segmentation models, namely, FDA~\cite{yang2020fda}, ADVENT~\cite{vu2019advent} and CyCADA~\cite{hoffman2018cycada}. The larger the number of blended images, the lower the improvement in Sim-to-Real performance. As a result, we just experimented with 40 blending images. If there is an improvement with this setting, it is inevitable with less blending quantity. 
The results are shown in Table~\ref{tab1}, indicating that our methods achieve higher mean IoU (mIoU) and mean Accuracy (mAcc) in terms of Sim-to-Real performance. Compared to the original settings of these models, although the increase in the number of blended images weakens the performance improvement, it still increases 4.96$\%$--9.85$\%$ in mIoU and 3.62$\%$--7.13$\%$ in mAcc, respectively.

\begin{table}[t!]
\begin{center}
\caption{Performances of SOTA domain adaptation methods adopting our IRB strategy. TS, BG, GL EP, UV mean train step, background, glottis, epiglottis, uvula, respectively.}
\label{tab1}
\resizebox{0.43\textwidth}{!}{
\renewcommand\arraystretch{1.3}
\begin{tabular}{ccccccc}
\toprule  %添加表格头部粗线
\multirow{2}{*}{Method}	&	\multirow{2}{*}{TS}	 	&	\multicolumn{5}{c}{\makecell*[c]{mIoU}} 				\\ \cline{3-7} 
	&	    	 &	BG	&	GL	&	EP	&	UV	&	mIoU	\\  
\cline{1-2} \cline{3-7}  
	&	40-r 	&	94.990 	&	72.030 	&	54.540 	&	65.660 	&	71.805 		\\
FDA~\cite{yang2020fda}	&	40-253	&	  96.620 	&	77.280 	&	74.260 	&	63.890 	&	78.013 	 	\\
	&	40-235		&	96.490 	&	78.440 	&	73.980 	&	66.590 	&	\textbf{78.875} 		\\
\cline{1-2} \cline{3-7} 
	&	40-r 	    	&	96.147 	&	80.790 	&	65.060 	&	67.900 	&	77.474 		 	\\
Advant~\cite{vu2019advent} &	40-253		    	&	97.100 	&	84.060 	&	79.330 	&	64.790 	&	\textbf{81.320} 			\\
	&	40-235	    	&	96.600 	&	78.290 	&	76.630 	&	67.460 	&	79.745 	\\
\cline{1-2} \cline{3-7} 
	&	40-r 		    	&	95.100 	&	72.930 	&	62.830 	&	59.080 	&	72.485 \\
Cycada~\cite{hoffman2018cycada}	&	40-253		    	&	94.640 	&	75.320 	&	72.380 	&	64.890 	&	\textbf{76.808} 		\\
	&	40-235		    	&	96.040 	&	69.790 	&	62.290 	&	63.620 	&	72.935 	\\

\bottomrule %添加表格底部粗线
\end{tabular}}
\vspace{-2em}
\end{center}
\end{table}

\section{Conclusion and Future Work}
In this work, we propose an image segmentation method targeting oropharyngeal organs with domain adaptive Sim-to-Real. By sorting the segmentation results between classes, IoU-Ranking Blend (IRB) provides a more appropriate blending strategy to maximize the potential of a limited number of blended images. Besides, we introduce the style-transfer method to further reduce the differences between datasets. The experiment results show that our proposed method improves the performance of the existing domain adaptive segmentation models.
In the future, we will deploy the pre-trained model into the software system of the intubation robot. EndoBot (Fig. \ref{Figure2}-D) will adjust the operating posture in real-time according to the segmentation results of the endoscopic transmission images.

%\addtolength{\textheight}{-12cm} 

%%%%%%%%%%%%%%%%%%%%%%%%%%%%%%%%%%%%%%%%%%%%%%%%%%%%%%%%%%%%%%%%%%%%%%%%%%%%%%%%

%%%%%%%%%%%%%%%%%%%%%%%%%%%%%%%%%%%%%%%%%%%%%%%%%%%%%%%%%%%%%%%%%%%%%%%%%%%%%%%%

\bibliographystyle{IEEEtran}
\bibliography{ref}
\end{document}